\documentclass{aastex}

\newif\ifEmulate
\def\Figure#1{\ifEmulate{#1}\fi}
\usepackage{emulateapj5} \Emulatetrue

\ifEmulate{\submitted{The Astrophysical Journal, 000:L000-L000, 2004 May 1}}\fi
 \shorttitle{The Bridge from White Dwarfs to M stars}
 \shortauthors{Smol\v{c}i\'{c} et al.}

\begin{document}

\title{A Second Stellar Color Locus: a Bridge from White Dwarfs to M stars}

\author{
Vernesa Smol\v{c}i\'{c}\altaffilmark{1,2},
\v{Z}eljko Ivezi\'{c}\altaffilmark{1,3},
Gillian R. Knapp\altaffilmark{1},
Robert H. Lupton\altaffilmark{1},
Kre\v{s}imir Pavlovski\altaffilmark{2},
Sa\v{s}a Iliji\'{c}\altaffilmark{4},
David Schlegel\altaffilmark{1},
J. Allyn Smith\altaffilmark{5},
Peregrine M. McGehee\altaffilmark{6},
Nicole M. Silvestri\altaffilmark{7},
Suzanne L. Hawley\altaffilmark{7},
Constance Rockosi\altaffilmark{7},
James E. Gunn \altaffilmark{1},
Michael A. Strauss\altaffilmark{1}
Xiaohui Fan\altaffilmark{8},
Daniel Eisenstein\altaffilmark{8},
Hugh Harris\altaffilmark{9}
}

\newcounter{address}
\setcounter{address}{1} \altaffiltext{\theaddress}{Princeton University Observatory, 
            Princeton, NJ 08544 \label{Princeton}}
\addtocounter{address}{1}
\altaffiltext{\theaddress}{
University of Zagreb, Dept. of Physics, Bijeni\v{c}ka cesta 32,
            10000 Zagreb, Croatia \label{Zagreb1}}
\addtocounter{address}{1}
\altaffiltext{\theaddress}{
H.N. Russell Fellow, on leave from the University of Washington \label{HNR}}
\addtocounter{address}{1}
\altaffiltext{\theaddress}{
Faculty of Electrical Engineering and Computing, Dept. of Applied Physics,
            Unska 3, 10000 Zagreb, Croatia \label{Zagreb2}}
\addtocounter{address}{1}
\altaffiltext{\theaddress}{
Department of Physics \& Astronomy, P.O. Box 3905, University of Wyoming,
            Laramie, WY 82071-3905 \label{Wyoming}}
\addtocounter{address}{1}
\altaffiltext{\theaddress}{
Los Alamos National Laboratory, MS H820, Los Alamos, NM 87545;
            also at New Mexico State University, Department of Astronomy, P.O. Box
            30001, Dept 4500, Las Cruces, NM 88003 \label{NewMexico}}
\addtocounter{address}{1}
\altaffiltext{\theaddress}{
University of Washington, Dept. of Astronomy,
            Box 351580, Seattle, WA 98195 \label{UW}}
\addtocounter{address}{1}
\altaffiltext{\theaddress}{Steward Observatory,
933 N. Cherry Ave., Tucson, AZ 85721\label{Steward}}
\addtocounter{address}{1}
\altaffiltext{\theaddress}{U.S. Naval Observatory,
Flagstaff Station, P.O. Box 1149, Flagstaff, AZ 86002\label{USNavy}}

\begin{abstract}
We report the discovery of a locus of stars in the SDSS 
$g-r$ vs. $u-g$ color-color diagram that connects the colors 
of white dwarfs and M dwarfs. While its contrast with respect 
to the main stellar locus is only $\sim$1:2300, this previously 
unrecognized feature includes 863 stars from the SDSS Data 
Release 1. The position and shape of the feature are in good 
agreement with predictions of a simple binary star model that 
consists of a white dwarf and an M dwarf, with the components' 
luminosity ratio controlling the position along this binary 
system locus. SDSS DR1 spectra for 47 of these objects strongly 
support this model. The absolute magnitude--color distribution 
inferred for the white dwarf component is in good agreement 
with the models of Bergeron et al. (1995).
\end{abstract}

\section{Introduction}

Modern large-scale accurate photometric surveys offer an unprecedented view of 
stellar populations. Here we discuss a population of unresolved binary stars
which account for fewer than 10$^{-3}$ of stars detected by the Sloan Digital
Sky Survey (York et al. 2000). Despite this low occurance frequency, the sample 
presented here is sufficiently large ($\sim$1000 stars) to characterize their 
broad-band optical properties.

\subsection{ Sloan Digital Sky Survey}

The Sloan Digital Sky Survey (SDSS; Abazajian et al.~2003, and references
therein) is revolutionizing stellar astronomy by providing homogeneous and deep
($r < 22.5$) photometry in five passbands ($u$, $g$, $r$, $i$, and $z$; 
Fukugita et al. 1996, Gunn et al. 1998, Hogg et al. 2001, Smith et al. 2002), 
accurate to 0.02 mag (Ivezi\'{c} et al.~2003). Ultimately, up to 10,000 deg$^2$ 
of sky in the Northern Galactic Cap will be surveyed. The survey sky coverage 
will result in photometric measurements for over 100 million stars and a similar 
number of galaxies. Astrometric positions are accurate to better than 0.1 arcsec 
per coordinate (rms) for point sources with $r<20.5^m$ (Pier et al.~2003), and 
the morphological information from the images allows robust star-galaxy separation 
to $r \sim$ 21.5$^m$ (Lupton et al.~2003).

Here we report the results of a color-based search for binary stars
in the recent SDSS Data Release 1 (see www.sdss.org), which includes
53 million unique objects detected in 2099 deg$^2$ of sky.

\subsection{     The Stellar Locus in the SDSS Photometric System     }

The effective temperature is the dominant parameter that determines the position
of the majority of stars in optical color-color diagrams constructed with
broad-band filters (Lenz et al. 1998, and references therein). The effective
temperature range results in a well-defined stellar locus in color-color diagrams
(for more details see  Finlator et al. 2000, and references therein).

\section{               The Locus of Binary Stars                 }

An unresolved binary star may have colors that place it either inside or
outside the locus. If the luminosity of one star is much greater than that
of the other, the more luminous star determines the system colors.
However, even if the luminosities are comparable, the system color may
still fall close to the locus of single stars in color-color diagrams where the
locus resembles a straight line (e.g. the \hbox{$i-z$}\ vs. \hbox{{$r-i$}}\ diagram).
Thus, to select unresolved binary systems by their colors, the most promising diagrams 
are those where the locus is curved, such as the \hbox{$g-r$}\ vs. \hbox{$u-g$}\ and
\hbox{{$r-i$}}\ vs. \hbox{$g-r$}\ color-color diagrams. The curvature in these 
diagrams is the result of saturation of the \hbox{$u-g$}\ and \hbox{$g-r$}\ colors
due to the strong molecular absorption bands which first appear at type $\sim$M0.

Figure 1 shows the \hbox{$g-r$}\ vs. \hbox{$u-g$}\ color-color diagram for 
$\sim$1.99 million stars from the public SDSS Data Release 1 (DR1) database 
with $u<20.5$. This magnitude limit ensures that the accuracy of the $u$ 
magnitudes is better than 0.1 mag (for $u<18$, the photometric accuracy is 
0.02 mag, Ivezi\'{c} et al. 2003). The most prominent feature is the main stellar
locus. Due to the large number of stars in DR1,
we are able to demonstrate the existence of a second stellar locus,
clearly visible just above the main stellar locus. The number of stars
in this, previously unreported feature, is a factor of $\sim$2,300 smaller than in the main
locus (using color cuts listed in the next Section). Thus, accurate multi-band
photometry ($u$ band in particular) for a sufficiently large number of stars
was required to detect such a low-contrast feature.

The second stellar locus is consistent with binary systems than include
an M dwarf and a white dwarf. We demonstrate that this simple model
provides a satisfactory explanation for the position of the second stellar
locus, hereafter the ``bridge'' (from M dwarfs to white dwarfs). We also show
that the available SDSS spectra for a subsample of bridge stars support this
interpretation.

\Figure{
 \centerline{\includegraphics[width=\hsize,clip]{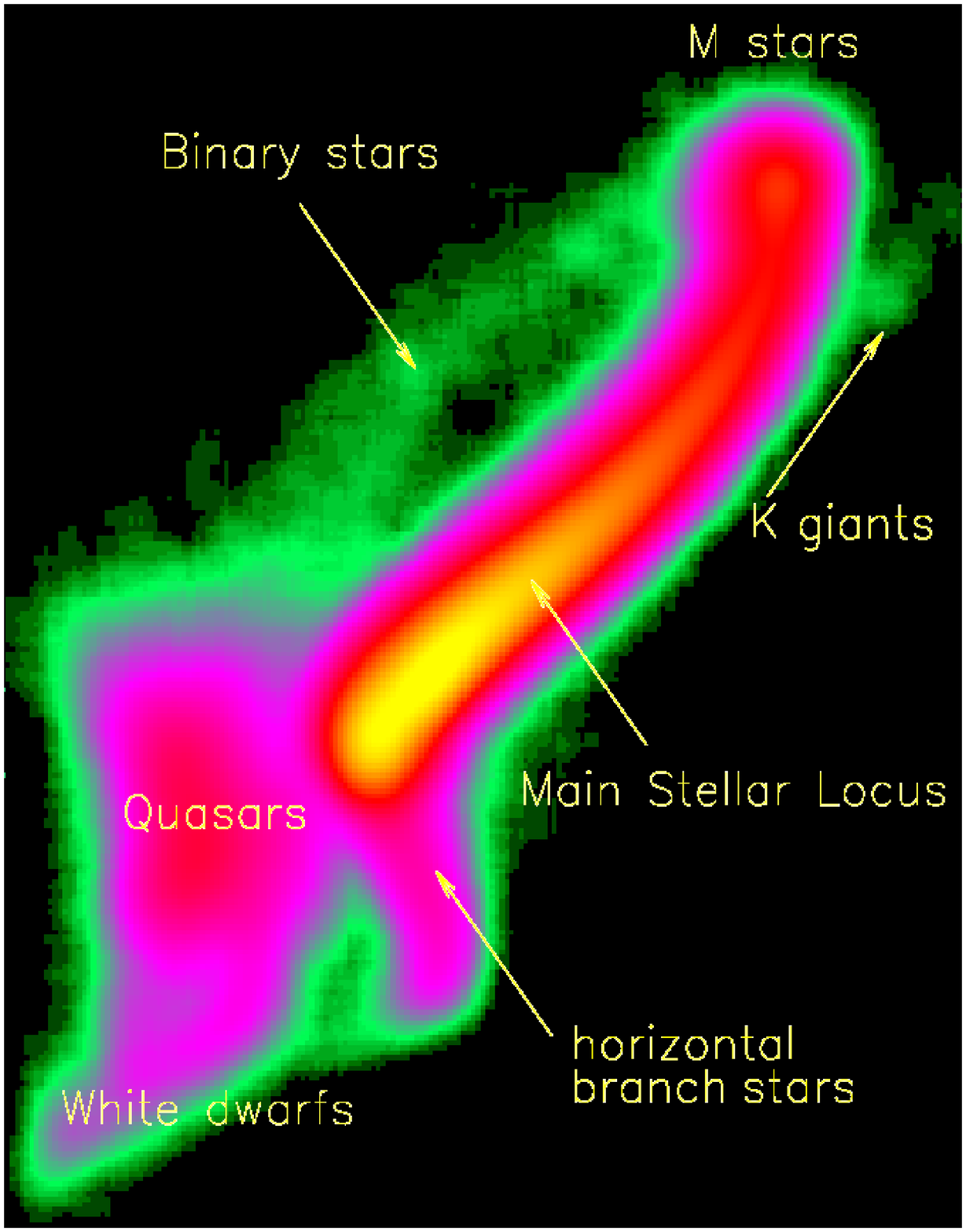}}
\figcaption[]{The number density, displayed on a logarithmic scale, of $\sim$1.99 
million stars with $u<20.5$ from SDSS Data Release 1 in the \hbox{$g-r$}\ vs. 
\hbox{$u-g$}\ color-color diagram (increasing from green to red to yellow; 
the color axes increase towards upper right and the color ranges are -1 to 3, 
and -1 to 2, for \hbox{$u-g$}\ and \hbox{$g-r$}, respectively). The most prominent 
features are the main stellar locus and the clump of low-redshift quasars, as marked. 
Other notable features include the locus of white dwarfs, horizontal branch stars 
(also including blue stragglers and RR Lyrae stars), and low-metallicity K giants. 
The fainter feature colored green, above and to the left of the main locus, 
is the locus of $\sim$1,000 binary stars. The properties of this locus are consistent 
with a distribution of M dwarf -- white dwarf pairs with varying luminosity ratio. 
The root-mean-scatter of stars about this locus is only $\sim$0.1 mag.
\label{ugrCMD}}
}

\section{              M dwarf -- White Dwarf Model             }

The bridge of stars in the \hbox{$g-r$}\ vs. \hbox{$u-g$}\ color-color diagram appears
to connect the positions of M stars and hot blue stars, with MK spectral
type around $B$. In order to produce a locus of stars that is not
coincident with the main stellar locus, the luminosities of the two
components must be comparable. The possibilities are an M dwarf -- white
dwarf pair, or an M giant -- blue giant/supergiant pair. The latter systems
cannot dominate the sample because the sky density of the selected stars
(0.40 deg$^{-2}$) is too high, given the faint magnitudes and high latitudes
probed by SDSS (see Majewski et al. 2004 for a nearly complete census of
M giants in the Galaxy).

We generate model colors for binary systems by assuming SDSS colors for
single M dwarf and white dwarf stars, and parametrize the system colors by
the luminosity ratio of the two components (in practice, we use the $r$ band
flux fractions). For the M dwarf we adopt $u-g = 2.6$, $g-r=1.4$,
$r-i = 2 (i-z)$, and $i-z=0.3$ to 0.75, with a step of 0.05, and for
the white dwarf $u-g=0.2$, $g-r=-0.2$, $r-i=0$, $i-z=0$. For more details
about M dwarfs and white dwarfs discovered by SDSS see Hawley et al. (2002),
Harris et al. (2003), Raymond et al. (2003) and Kleinman et al. (2004). 
The model predictions are compared to the data in Figure 2, where the dots 
represent 863 ``DR1 bridge stars'', selected by requiring $u < 20.5$, 
\hbox{$u-g$}$<$2, \hbox{$g-r$}$>$0.3, \hbox{$r-i$}$>0.7$, and that processing 
flags BRIGHT, SATUR and BLENDED are not set (the flag requirement selects 
unique unsaturated objects, see Abazajian et al. 2003). We corrected all 
colors for the interstellar reddening using the maps from Schlegel, Finkbeiner 
\& Davis (1998); typical corrections at the high galactic latitudes considered 
here are $<0.05$ mag.

\Figure{
 \centerline{\includegraphics[width=\hsize,clip]{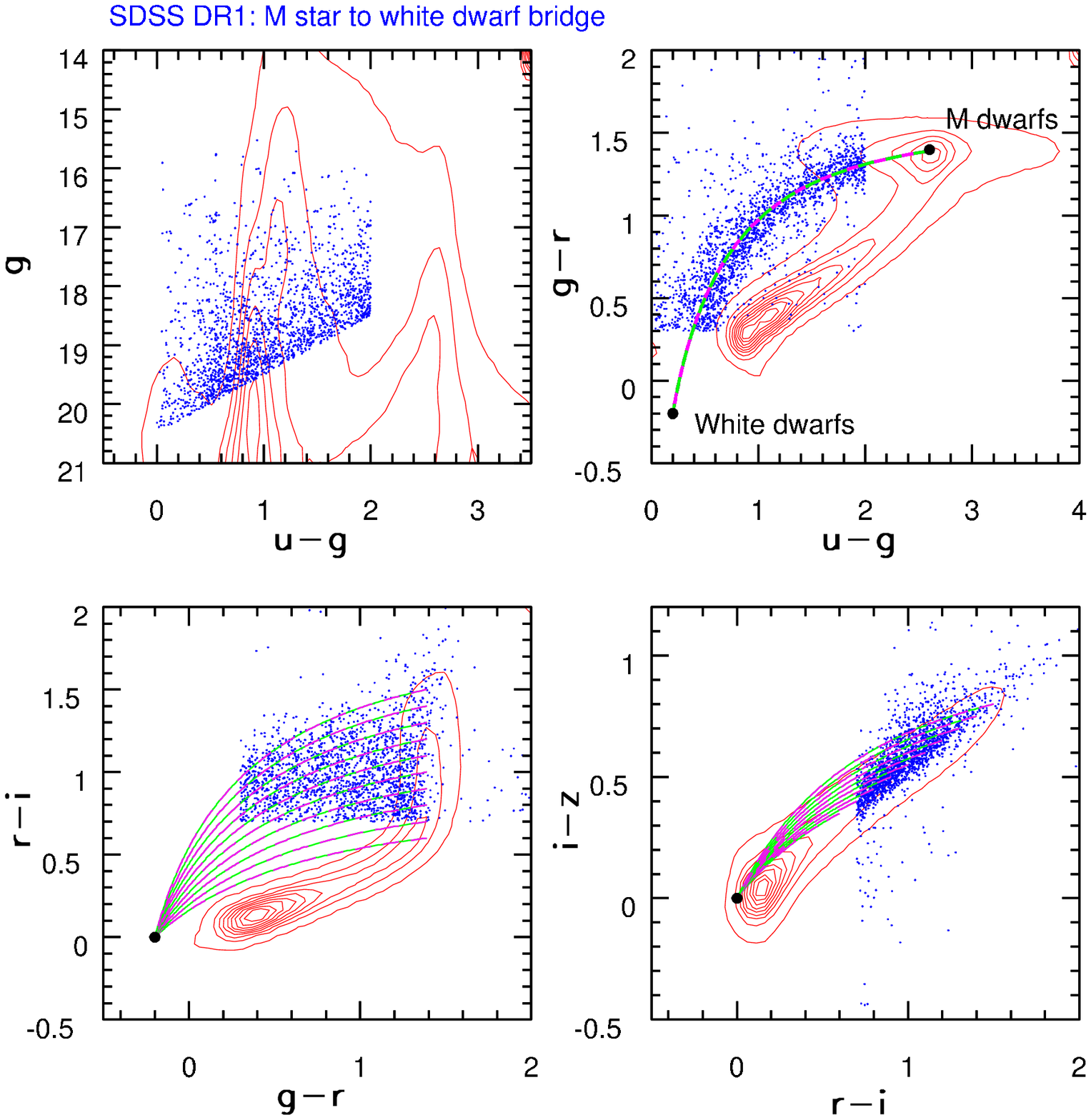}}
\figcaption[]{The comparison of a simple M dwarf -- white dwarf pair model 
with the data. A representative distribution of all stars is shown by linearly 
spaced isopleths. 863 stars from SDSS Data Release 1 selected by requiring 
$u < 20.5, \hbox{$u-g$}<2, \hbox{$g-r$}>0.3$ and $\hbox{{$r-i$}}>0.7$
are shown by dots. The model predictions are shown by lines, where each
line corresponds to different \hbox{{$r-i$}}\ and \hbox{$i-z$}\ colors assumed 
for the M dwarf. The position along the line depends on the luminosity ratio
of the two components. There is only one line in the \hbox{$g-r$}\ vs. \hbox{$u-g$}\
diagram (top right) because all M dwarfs have practically the same \hbox{$u-g$}\ and 
\hbox{$g-r$}\ colors (Finlator et al. 2000, Hawley et al. 2002). The observed data 
scatter around this line is presumably due to a distribution of white dwarf 
colors, photometric errors, and sample contamination by other types of source.
\label{CMD}}
}

\newpage
The agreement between this simple binary star model and the data is satisfactory.
In particular, the model track closely follows the distribution of bridge stars
in the \hbox{$g-r$}\ vs. \hbox{$u-g$}\ color-color diagram, and reproduces the observed range of
colors in other diagrams. A noteworthy point is that the implied contribution of
the white dwarf to the total $r$ band flux is at most 50\% for practically all
stars from the ``DR1 bridge'' sample (for equal $r$ band flux contributions, the
model predicts $u-g=0.40$ and $g-r=0.33$).

\Figure{
 \centerline{\includegraphics[width=\hsize,clip]{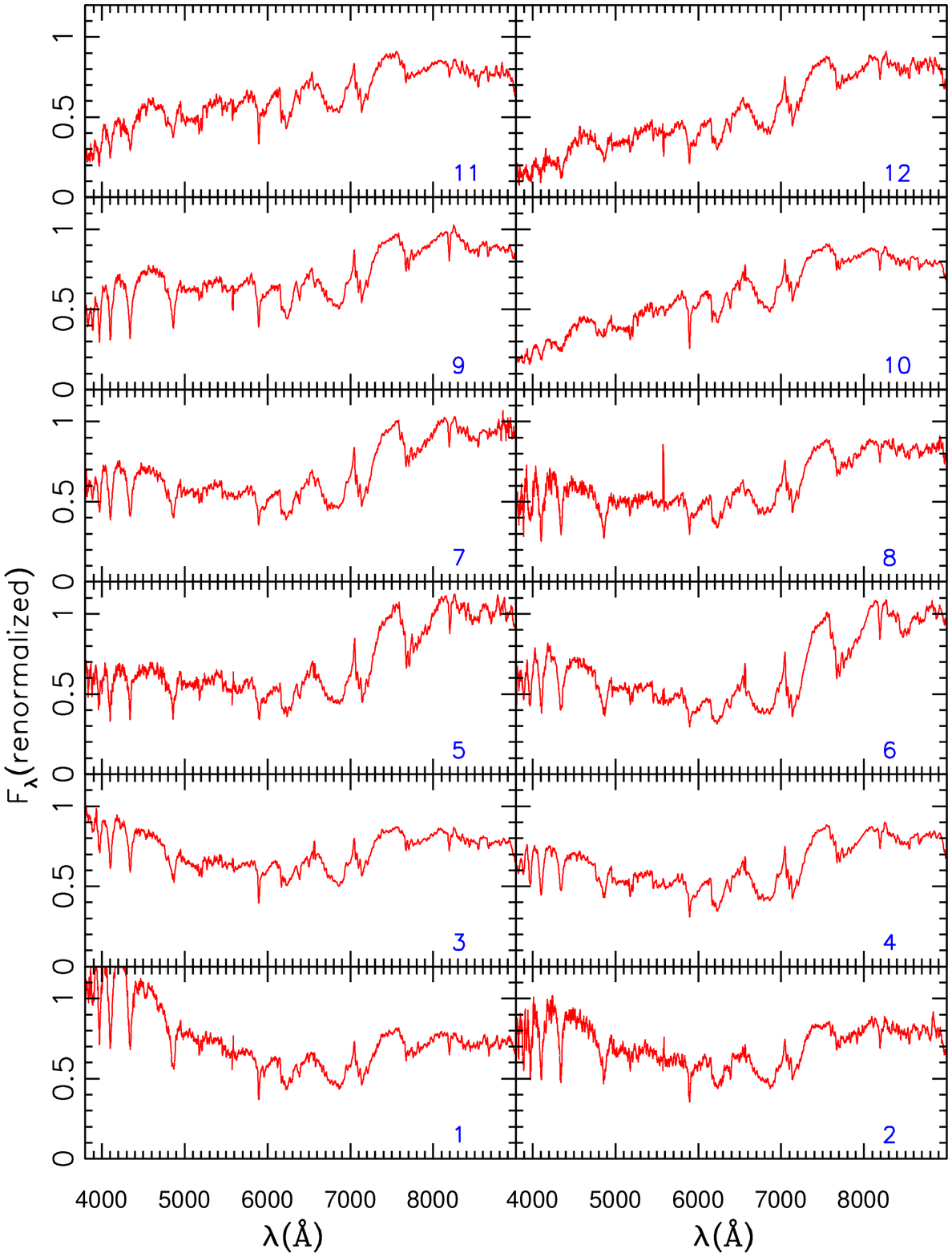}}
\figcaption[]{SDSS spectra for a subsample of stars whose SDSS colors are consistent 
with an M dwarf -- white dwarf binary system. In all the shown systems the red
half of the spectrum is typical of an M dwarf, while the blue half is consistent
with a white dwarf spectrum. Spectra are approximately ordered by the M dwarf 
to white dwarf flux ratio in the $r$ band. Note prominent H$\alpha$ emission for 
stars 3 and 6. The feature at $\sim$5,577\AA\ is due to the night sky.
\label{spectra}}
}

\section{  SDSS Spectra  }

While the close agreement between the data and model predictions supports
the hypothesis that the bridge stars are dominated by M dwarf -- white dwarf
pairs, further confirmation can be gained by examining the available SDSS spectra.
Stars are selected by various criteria for SDSS spectroscopic observations,
and we postpone a detailed analysis of the selection statistics to a forthcoming
publication. Here we simply report the results of a visual examination of 47 ``bridge'' 
stars from the DR1 sample for which SDSS spectra are available. The spectra are obtained
through 3$^{\arcsec}$ fibers, and span the wavelength range 3800--9200\AA,
with a spectral resolution of $\lambda/\Delta \lambda \sim 1800$.

Out of 863 stars in the sample, spectra are available for 47. The visual
inspection of spectra indicates that 45 are consistent with an M dwarf--white
dwarf interpretation (the remaining 2 are G stars; both are close to the
color-selection boundary, and one belongs to a complex blended source). We display 
a representative sample of spectra in Figure 3. A preliminary comparison with the 
M dwarf spectral sequence (Hawley et al. 2002) indicates that M dwarfs in the 
binary systems discussed here are dominated by types M5 and earlier, as is the 
case for single M dwarfs in this magnitude-limited sample. This conclusion is 
also supported by the distribution of their \hbox{{$r-i$}}\ and \hbox{$i-z$}\ 
colors (the median \hbox{{$r-i$}}\ color is $\sim$1.0, see the lower left panel 
in Figure 2). 

We visually compared all spectra to the atlas of white dwarf spectra
by Wesemael et al. (1993). About half belong to the DA class, and
about one third can be tentatively classified as subdwarfs. Other classes
that are probably present in the sample include DB, DZ and DQ. 
We note that SDSS spectra are of sufficient quality to allow determination 
of the white dwarf temperature and the M dwarf chromospheric
activity (via H$\alpha$ emission), as demonstrated by Raymond et al. (2003).
Such an analysis will be presented in a forthcoming publication.

\Figure{
 \centerline{\includegraphics[width=\hsize,clip]{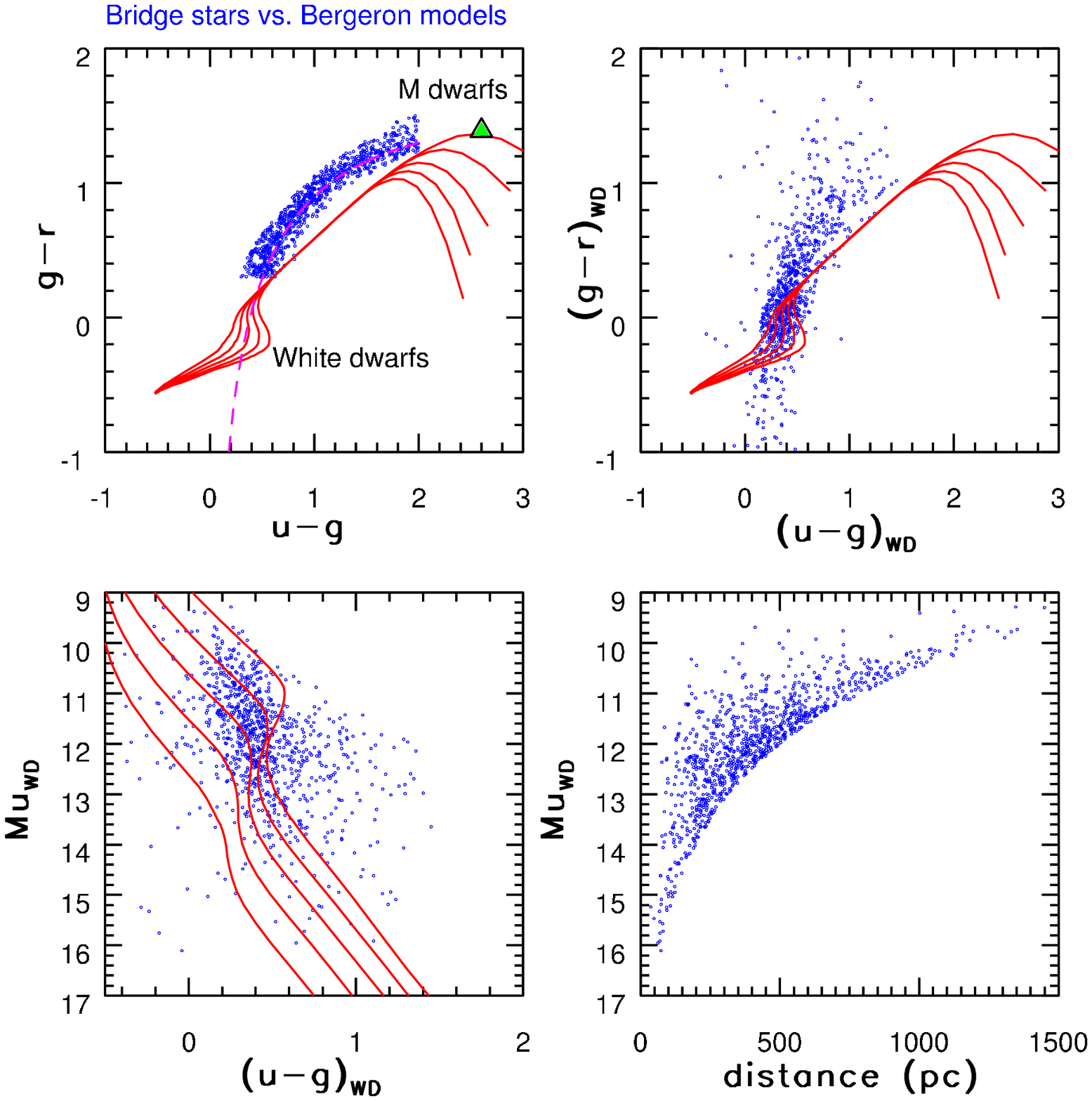}}
\figcaption[]{The comparison of data with Bergeron et al. (1995) white 
dwarf models. The symbols in the top left panel display the $g-r$ vs. 
$u-g$ color distribution for bridge stars, and the lines are white dwarf 
models with log(g)=(7, 7.5, 8, 8.5, 9).
For a given $g-r$ color, the models with larger log(g) have bluer $u-g$ color
(the temperature range is from 1500 K to 10$^5$ K). The top right panel is 
analogous to the top left panel, except that the symbols show the color
distribution for the white dwarf component. The bottom left panel compares
the color-magnitude distribution for the white dwarf component with the 
model predictions (for a given $u-g$ color, log(g) decreases with the luminosity).
The white dwarf absolute magnitude -- distance distribution is shown in the 
bottom right panel.
\label{bergeron}}
}

\section { Comparison with White Dwarf Models  }

The models discussed in Section 3 indicate that the white dwarf contribution to 
the \hbox{$i$}\ and \hbox{$z$}\ band fluxes is practically negligible. Hence, 
the absolute \hbox{$i$}\ band magnitude for the M dwarf component, $M_i$,  
and therefore distances, can be 
obtained using the $M_i$ vs. \hbox{$i-z$}\ color-magnitude relation from Hawley 
et al. (2002). With an estimate for distance, the absolute $u$ band magnitude 
for the white dwarf component can be determined, and compared to model cooling
curves. Two additional parameters that can be derived from the data are the
$u-g$ color for the white dwarf component, and the components' $r$ band 
flux (or luminosity) ratio. 
In this analysis, we further constrain the sources to be very close 
(0.15 mag) to the ``bridge'' by requiring $P_2 < 0.5\,P_1^2 + 0.15$ and 
$P_2 > 0.2\,P_1^2 - 0.15$, where $P_1 = -0.5\,(u-g) - (g-r) + 1.5$ and 
$P_2 = 0.7\,(u-g) - (g-r) + 0.2$.

Using this approach we find a median $M_i\sim9$ for the M dwarf component,
with a root-mean-scatter of 1 mag. The corresponding median distance is $\sim$400 pc. 
The derived white dwarf parameters are compared to models by Bergeron et al. 
(1995) in Fig. 4. As shown in the top right panel, the estimated white dwarf colors
agree well with the model predictions. Similarly, fairly good agreement 
is obtained for the luminosity--color distribution displayed in the bottom
left panel. Models with 7$<$log(g)$<$8.5 bracket the majority of data points,
in agreement with the analysis of isolated white dwarfs with 
SDSS spectra (Kleinman et al. 2004). About 20\% of data points have $u-g$ color, 
for a given absolute magnitude, too red by about 0.5 mag (or equivalently, 
for a given $u-g$ color, absolute magnitude is too bright by $\sim$2 mag). 
This discrepancy could be due to sample contamination by other types of source. 

The white dwarf absolute magnitude -- distance distribution is shown in the 
bottom right panel. Since the sample presented here is an unbiased, $u$ flux-limited 
sample (with the adopted $u$ magnitude limit, the other four SDSS magnitudes for 
all stars in the sample are comfortably brighter than the corresponding SDSS 
completeness limits), it would be straightforward to determine the white dwarf 
luminosity function and the number density (assuming that the components are 
not strongly interacting). However, the unresolved binary stars discussed here 
are heavily biased towards systems with components that have similar luminosities, 
and it is not trivial to account for this effect. Such a detailed analysis of 
M dwarf and white dwarf luminosity functions in unresolved binary systems
will be presented in a forthcoming publication.

\section{  Conclusions  }

The accurate multi-band SDSS photometry for a large number of stars allowed
detection of a new feature in the broad-band optical color-color diagrams: 
a ``bridge'' of stars, well-separated from the main stellar locus, connects the 
positions of M dwarfs and white dwarfs. The bridge characteristics are consistent 
with a binary system than includes an M dwarf and a white dwarf, with the system's 
position on the bridge determined by the components' luminosity ratio. This 
conclusion is strongly supported by SDSS spectra for 47 such systems. 

The distance to these systems can be estimated in a straightforward way 
because a  photometric parallax relation for M dwarfs can be applied to 
$i$ and $z$ band measurements, where the contribution from the white dwarf
is negligible. With a known system distance, the white dwarf luminosity-color
distributions can be determined and compared to models. We find that models
by Bergeron et al. (1995) are in good agreement with the data. 

This work analyzed only about a quarter of the data that will be obtained
by the SDSS. Thus, the color selection method presented here will eventually
yield $\sim$4,000 unresolved M dwarf-- white dwarf binary systems.

\vskip 0.3in
{\it Acknowledgements}

We thank Princeton University for a generous support of this research. 

Funding for the creation and distribution of the SDSS Archive has been provided by
the Alfred P. Sloan Foundation, the Participating Institutions, the National Aeronautics
and Space Administration, the National Science Foundation, the U.S. Department of Energy,
the Japanese Monbukagakusho, and the Max Planck Society. 
The SDSS Web site is http://www.sdss.org/.

\ifEmulate\end{document}\fi

\end{document}